\documentclass[rnote]{aa}
\usepackage{amsmath}
\usepackage{graphicx}
\usepackage{natbib}
\usepackage{txfonts}
\bibpunct{(}{)}{;}{a}{}{,}
\begin{document}
%
\title{The applicability of the wind compression model}

\author{Z. Carikov\'{a} \and A. Skopal}

\institute{Astronomical Institute, Slovak Academy of Sciences,
          059~60 Tatransk\'{a} Lomnica, Slovakia\\
		  \email{zcarikova@ta3.sk}}
		
\date{Received / Accepted}

\abstract
{Compression of the stellar winds from rapidly rotating hot stars
is described by the wind compression model. However, it was also
shown that rapid rotation leads to rotational distortion
of the stellar surface, resulting in the appearance of non-radial
forces acting against the wind compression. 
In this note we justify the wind compression model for moderately
rotating white dwarfs and slowly rotating giants. The former could 
be conducive to understanding density/ionization structure of 
the mass outflow from symbiotic stars and novae, while the 
latter can represent an effective mass-transfer mode in the wide 
interacting binaries.}
\maketitle
\section{Introduction}

Rotation of the hot stars with radiation-driven winds leads to
compression of the outflowing material towards the equatorial
regions. The mechanism is described by the wind compression
model developed by \citet{bjorkcass93}.
Because of conservation of the angular momentum vector,
each particle launched at the surface of the star has to move
in its own orbital plane, which is perpendicular to this vector.
The streamlines of the gas from both hemispheres of
a rotating wind are bent towards the equatorial
plane. As a result the wind is compressed towards equatorial
regions of the star.
Two forms of the equatorial density enhancements can be recognized:
(i) if the flow from the hemispheres has a supersonic component
of the velocity perpendicular to the equatorial plane, a shock
zone or disk will develop and we get the wind-compressed disk
(WCD) model elaborated by \citet{bjorkcass93}, and
(ii) if the streamlines do not cross the equatorial plane of
the star we get the wind-compressed zone (WCZ) model described
by \citet{igncassbjork96}.

However, following studies on the mass loss from the {\em oblate}
stellar surface, caused by a rapid rotation of the star, showed
the presence of inverse effects inhibiting the equatorial wind
compression.
\citet{craowo95} investigated the effect of oblateness
and gravity darkening on the radiation driving in winds from
rapidly rotating Be stars. They revealed the presence of
non-radial radiative driving, directed away from the equatorial
plane and acting against the sense of rotation. These effects
thus weaken the equatorward wind compression compared to wind
models calculated for spherical stars.
Inhibition of the WCD formation was also confirmed by
radiation-hydrodynamical simulations of \citet{owcrga96},
who showed that these non-radial forces can lead to an effective
{\em suppression} of the equatorward flow needed to form a WCD.

An interesting effect of the rotational distortion of the star
is gravity darkening. According to the von Zeipel theorem, the
surface radiative flux is proportional to the local
effective gravity, $F \propto g_{\rm eff}$, which is reduced at
the equator. Since the poles of the star have a higher
$g_{\rm eff}$, they also have a higher temperature
$T_{\rm eff}$ ($T_{\rm eff} \propto g_{\rm eff}^{1/4}$) and
thus higher brightness $F$. On the other hand, the equatorial
regions of the star have a lower $g_{\rm eff}$ and thus also
a lower brightness $F$ (called gravity darkening).
As a result, the gravity darkening reduces the mass loss rate
from equatorial regions and strengthens it toward the relatively
bright poles of rapidly rotating Be stars
\citep{owcrga96,owgacr98}.

In this contribution we justify applicability of the wind
compression model to other rotating stellar objects, whose
observational phenomena could be understood within this model.
In particular, to (i) white dwarfs (WDs) in symbiotic
binaries and cataclysmic variables, whose presumable moderate
rotation could influence the shaping of their mass outflow during
outbursts, and (ii) slowly rotating red giants (RGs) in wide
interacting binaries, where the wind compression towards
the equatorial plane can significantly enhance the accretion
from the wind of the giant by its compact companion.
In Sect.~2 we introduce distortion of the stellar
surface due to rotation; Sect.~3 gives examples of such
flattening for a rapidly rotating Be star, moderately rotating
WD, and slowly rotating RG. A brief discussion and
conclusion are found in Sects.~4 and 5, respectively.

\section{Rotational distortion of the stellar surface}

Rotation of a star leads to a distortion of its surface. 
The shape of the star, i.e. the latitude dependence of the 
stellar radius $R(\theta)$, where $\theta$ is the stellar 
co-latitude with $\theta = 0$ at the pole of the star, is determined 
by the equipotential surfaces. For the rigid rotation, the 
ratio $R(\theta)$ to its value at the equator, $R_{\rm eq}$, 
can be expressed as \citep[see][]{kraus06} 
\begin{equation}
  \frac{R(\theta)}{R_{\rm eq}} =
  2\frac{\sqrt{2+\omega^{2}}}{\sqrt{3}\omega \sin{\theta}}
  \sin{\left\{ \frac{1}{3}
  \arcsin \left(\frac{3\sqrt{3}\omega\sin\theta}
  {\left(2+\omega^{2} \right)^{3/2}} \right)
  \right\}},
\label{eq:shape}
\end{equation}
where the limit for $\theta=0$ yields the ratio
\begin{equation}
  \frac{R_{\rm pole}}{R_{\rm eq}} = 
  \left(1+\frac{1}{2}\omega^{2}\right)^{-1},
\label{eq:rr}
\end{equation}
where the polar radius $R_{\rm pole} = R(\theta=0)$, 
$\omega = v_{\rm rot}/v_{\rm crit}$, and $v_{\rm rot}$ 
is the equatorial rotation velocity of the star. The critical 
velocity $v_{\rm crit}$ can be written as 
\begin{equation}
  v_{\rm crit} = \sqrt{\frac{GM_{\rm eff}}{R_{\rm eq}^{\rm max}}} 
               = \sqrt{\frac{GM_{\star}(1-\Gamma_{\rm e})}
                            {R_{\rm eq}^{\rm max}}},
\label{eq:vcrite}
\end{equation}
where $G$ is the gravitational constant, $R_{\rm eq}^{\rm max}$ 
is the maximum equatorial radius when the star is rotating at 
the critical velocity (i.e. for $\omega=1$), and $M_{\rm eff}$ is 
the effective stellar mass, i.e. the stellar mass reduced by 
the effects of radiation pressure due to electron scattering. 
It is defined as $M_{\rm eff} = M_{\star}(1-\Gamma_{\rm e})$, 
where $M_{\star}$ is the mass of the star and
$\Gamma_{\rm e}$ is the Eddington factor. 
For a non-rotating star, $\Gamma_{\rm e}$ is the ratio of 
the stellar luminosity $L_{\star}$ to the Eddington luminosity 
$L_{\rm Edd}$, i.e. $\Gamma_{\rm e}=L_{\star}/L_{\rm Edd}$. 
The Eddington luminosity is given by 
\begin{equation}
  L_{\rm Edd} = \frac{4\pi cGM_{\star}}{\kappa_{\rm T}},
\end{equation}
where $c$ is the speed of the light, and $\kappa_{\rm T}$ is the 
electron scattering opacity per unit mass, which depends on 
the chemical composition of the wind and the degree of 
ionization. For example, 
$\kappa_{\rm T}\sim$ 0.30\,cm$^{2}$\,g$^{-1}$ for the winds
of hot stars \citep{lamcass99}. 

According to \citet{maemey2000} relation (\ref{eq:vcrite})
with a general $\Gamma$
(i.e. including total opacity, not only electron opacity)
is true if we assume that the brightness of
the rotating star is uniform over its surface, which is in
contradiction to the von Zeipel theorem.
In rotating stars the effective gravity and the flux vary
over the stellar surface (Sect.~1). Therefore, we have to
consider the local Eddington factors $\Gamma$ \citep{maeder99}.
\citet{maemey2000} showed that the correct Eddington factors
in a rotating star depend on the rotation. As a consequence, the
maximum possible stellar luminosity is reduced by rotation.
The critical velocity has a different
expression depending on the luminosity of the star.
If the Eddington factor $\Gamma < 0.639$, the critical velocity
is independent of the Eddington factor and is given by the classical
expression \citep[e.g.][]{puls+08}

\begin{equation}
  v_{\rm crit} = \sqrt{\frac{GM_{\star}}{R_{\rm eq}^{\rm max}}}
               = \sqrt{\frac{2GM_{\star}}{3R_{\rm pole}}}.
\label{eq:vcritc}
\end{equation}
Physically, this means that when the luminosity of the star
is sufficiently below the Eddington limit, the reduction
of the effective mass by the rotation is not so important
and the classical Eq. (\ref{eq:vcritc}) can be used.
For $\Gamma > 0.639$, the critical velocity is significantly
reduced by the proximity to the Eddington limit. When the
Eddington factor $\Gamma$ tends towards 1,
the critical velocity tends to 0
\citep[for more details see][]{maemey2000}.

\section{Examples of different types of rotating stars}

In this section we compare the oblateness of a rapidly
rotating Be star, rotating WD, and RG in symbiotic binaries.
Table~1 introduces the ratio of the polar radius to
the equatorial radius $R_{\rm pole}/R_{\rm eq}$
calculated according to Eq. (\ref{eq:rr})
for typical parameters of selected types of stars.

\subsection{Rapidly rotating Be star}

To study inhibition of the WCD formation by non-radial
line forces in rotating hot star winds, \citet{owcrga96}
chose model S350 by \citet{ocb94} as a representative Be star model:
$v_{\rm rot} = 350$ km\,s$^{-1}$,
$M_{\star} = 7.5$\,M$_{\sun}$,
$R_{\rm pole} = 4$\,R$_{\sun}$
(at the critical velocity $R_{\rm eq}^{\rm max} = 6$\,R$_{\sun}$),
and $L_{\star} = 2\,310$\,L$_{\sun}$.
Equation (\ref{eq:vcrite}) gives $v_{\rm crit} = 487$ km\,s$^{-1}$
(the Eddington factor $\Gamma_{\rm e} = 0.007$), while
the classical expression Eq. (\ref{eq:vcritc}) yields
$v_{\rm crit} = 488$ km\,s$^{-1}$.
Thus, according to Eq. (\ref{eq:rr}) the ratio of the polar to 
the equatorial radius $R_{\rm pole}/R_{\rm eq} \sim 0.794$.
As was shown by \citet{ocb94}, the wind particles emitted by such
a flattened stellar surface cannot form the WCD or WCZ.

\subsection{Moderately rotating WD}

Here we consider a typical WD in symbiotic binary with
$v_{\rm rot} = 300$ km\,s$^{-1}$ \citep{zcsk12},
$M_{\star} = 1$\,M$_{\sun}$ ,
$R_{\rm pole} = 0.01$\,R$_{\sun}$,
$L_{\star} = 1\,000$\,L$_{\sun}$.
Then $v_{\rm crit} = 3\,520$ km\,s$^{-1}$ ($\Gamma_{\rm e} = 0.023$)
or $v_{\rm crit} = 3\,560$ km\,s$^{-1}$
according to Eq. (\ref{eq:vcrite}) or Eq. (\ref{eq:vcritc}), respectively.
This corresponds to the ratio $R_{\rm pole}/R_{\rm eq} \sim 0.996$.
During active phases the luminosity of the WDs can increase
to $\sim 10^{4}$ L$_{\sun}$. However, the Eddington factor
$\Gamma_{\rm e}$ would be still lower than 0.639, so the classical
expression for the critical velocity given by Eq. (\ref{eq:vcritc})
can be used.
Thus, in the case of rotating WDs the effects of the stellar surface
distortion to inhibition of the wind compression to the equatorial
plane will probably be negligible.
An application of the WCZ model to moderately rotating
WDs in symbiotic binaries was recently suggested by \citet{zcsk12}.

\subsection{Slowly rotating RG}

According to \citet{zam+08}, the median $v_{\rm rot} \sin i$
for M0-M6 giants in symbiotic binaries is $\sim$ 8 km\,s$^{-1}$.
Thus we adopt
$v_{\rm rot} = 8$ km\,s$^{-1}$,
$M_{\star} = 1$\,M$_{\sun}$ ,
$R_{\rm pole} = 100$\,R$_{\sun}$ \citep[e.g.][]{bel+99},
$L_{\star} = 1\,600$\,L$_{\sun}$ \citep[e.g.][]{sk05}.
Then Eq. (\ref{eq:vcrite}) and Eq. (\ref{eq:vcritc}) yield
$v_{\rm crit} = 35$ km\,s$^{-1}$ ($\Gamma_{\rm e} = 0.037$)
and $36$ km\,s$^{-1}$, respectively.
According to Eq. (\ref{eq:rr}) the flattening ratio
$R_{\rm pole}/R_{\rm eq} \sim 0.975$.
This result suggests that the equatorward compression
of wind particles emitted by the surface of a slowly rotating RG
can be expected (see Sect.~4).

\begin{table}
\begin{center}
\caption{The ratio $R_{\rm pole}/R_{\rm eq}$ for a Be star,
WD, and RG and possible applicability of the WCZ model
(Sect.~3).}
\begin{tabular}{cccc}
\hline
\hline
 Star & $R_{\rm pole}/R_{\rm eq}$ & Applicability & Ref.\\
\hline
  Be       & 0.794 & No  & 1,2\\
  WD$^{a}$ & 0.996 & Yes & 3,4\\
  RG$^{b}$ & 0.975 & Yes & 4\\
\hline
\end{tabular}
\end{center}
\tablefoot{\\
1 -- \citet{craowo95}\\
2 -- \citet{owcrga96}\\
3 -- \citet{zcsk12}\\
4 -- this paper\\
$^{(a)}$ in symbiotic binaries\\
$^{(b)}$ normal RG in S-type symbiotic binaries}
\end{table}

\section{Discussion}

The above mentioned examples suggest that we can eliminate
the problem with the distortion of the stellar surface of 
rapidly rotating hot stars for compact objects (e.g. WDs), 
which cannot be easily deformed at moderate rotational 
velocities, and/or objects rotating slowly (e.g. RGs). 

Recently, we applied the wind compression model to the WDs
in symbiotic binaries \citep{zcsk12}. We calculated the WCZ 
formation of their enhanced winds during active 
phases and the corresponding ionization structure. Both 
the geometrical and ionization properties of our models were 
consistent with those derived from observations. In this way 
we justified the applicability of the wind compression 
model to WDs in symbiotic binaries. 

In the light of the above mentioned case, the observed 
non-spherical structure of the ejected material during the 
nova event 
\citep[see e.g. the recent summarization by][]{shore13} 
could also have an origin in the rotation of a WD, from whose
surface the accumulated material is ejected.
This possibility was investigated for the first time
by Ignace et al. (1996). They found that the wind compression
model may have relevance for slow novae, which tend to maintain
winds with low terminal speeds of a few hundred km\,s$^{-1}$
throughout their outbursts. Recently, the observed evolution
of the biconical ionization structure of fast novae
V339~Del ($v_{\infty} \sim$ 2\,700 km\,s$^{-1}$)
and RS~Oph ($v_{\infty} \sim$ 4\,000 km\,s$^{-1}$)
was also possible to interpret qualitatively in terms
of the WCZ model \citep[see][]{sk+14,sk15na2}.
However, a theoretical application of the WCZ model to the
fast nova mass outflow is needed to test this interpretation.

As the flattening ratio for RGs in symbiotic binaries is 
comparable with that of their WDs (Table~1), 
one can assume that the inhibition of the WCZ (or WCD) 
formation will also be negligible for the slowly rotating 
RGs. Here, the major astrophysical consequence is a possible 
increase in the wind mass transfer efficiency in wide 
interacting binaries containing an evolved star, which could be
an alternative to gravitationally focused wind accretion 
\citep[see][]{borro+09,moh+pod12}.
A particular application of the WCZ model to evolved giants
was introduced by Ignace et al. (1996), who calculated
a density contrast between the equator and pole as
a function of the rotation rate
$\omega = v_{\rm rot}/v_{\rm crit}$ and the parameter
$\beta$, which characterizes the wind acceleration.
For comparison, a slow wind of red giants with
$\beta = 2.5$ \citep{schroder85} and $\omega = 0.23$ (Sect.~3.3)
imply the density contrast of $\sim$ 9
\citep[see Fig.~3 for AGB stars in][]{igncassbjork96}, which
suggests a significant effect of the wind compression.
A detailed application of the WCZ model to stellar winds of
normal giants in S-type symbiotic stars is presented in our
following paper \citep[][submitted]{skzc14}.

\section{Conclusion}

Previous work of \citet{craowo95} and \citet{owcrga96}
demonstrated that a strong stellar surface distortion of
the rapidly rotating Be stars causes the appearance of
non-radial line forces and the effect of gravity darkening
that effectively suppress the equatorward flow needed to form
a WCD.

In this note we investigated the possibility of applying the wind
compression model to moderately rotating WDs in interacting
binaries, e.g. symbiotic stars, and to slowly rotating RGs in
wide interacting binaries.
We determined the rotational flattening ratio
$R_{\rm pole}/R_{\rm eq} \sim$ 0.794, 0.996, and 0.975
for parameters of a typical
Be star, WD, and RG in a symbiotic binary, respectively.
Applicability of the WCZ formation to WDs in symbiotic stars
was already demonstrated by \citet{zcsk12}.
Consequently, the very small distortion of slowly rotating RGs
in symbiotic stars, which is comparable to that of their WDs,
suggests that the wind compression model can also be applied
to the slowly rotating RGs in wide interacting binaries.

The first case can be used in modelling the structure of the
enhanced mass outflow (wind) from WDs during outbursts of
symbiotic stars and/or novae, while the second case can be
used to model a very efficient wind mass transfer from a giant
to its WD companion as indicated in wide interacting binaries.

\begin{acknowledgements}
The authors thank the anonymous referee for constructive comments.
Daniela Kor\u{c}\'akov\'a is thanked for discussions.
This project was supported by the Slovak Academy of Sciences 
under grant VEGA No.~2/0002/13.
\end{acknowledgements}


\begin{thebibliography}{}
%
\bibitem[Bjorkman \& Cassinelli (1993)]{bjorkcass93}
         Bjorkman, J. E., Cassinelli, J.P. 1993, ApJ, 409, 429
%
\bibitem[Carikov\'a \& Skopal (2012)]{zcsk12}
         Carikov\'a, Z., Skopal, A. 2012, A\&A, 548, A21
%
\bibitem[Cranmer \& Owocki (1995)]{craowo95}
         Cranmer, S. R., Owocki, S. P. 1995, ApJ, 440, 308
%
\bibitem[de Val-Borro et al. (2009)]{borro+09}
         de Val-Borro, M., Karovska, M., \& Sasselov, D. 
         2009, ApJ, 700, 1148
%
\bibitem[Ignace et al. (1996)]{igncassbjork96}
         Ignace, R., Cassinelli, J. P., Bjorkman, J. E. 
         1996, ApJ, 459, 671
%
\bibitem[Kraus (2006)]{kraus06}
         Kraus, M. 2006, A\&A, 456, 151
%
\bibitem[Lamers \& Cassinelli (1999)]{lamcass99}
         Lamers, H. J. G. L. M., Cassinelli, J. P. 
         1999, Introduction to stellar winds, 
         Cambridge University Press
%
\bibitem[Maeder (1999)]{maeder99}
         Maeder, A. 1999, A\&A, 347, 185
%
\bibitem[Maeder \& Meynet (2000)]{maemey2000}
         Maeder, A., Meynet, G. 2000, A\&A, 361, 159
%
\bibitem[Mohamed \& Podsiadlowski (2012)]{moh+pod12}
         Mohamed, S., \& Podsiadlowski, Ph. 
         2012, Baltic Astronomy, 21, 88
%
\bibitem[Owocki et al. (1994)]{ocb94}
         Owocki, S. P., Cranmer, S. R., Blondin, J. M. 
         1994, ApJ, 424, 887
%
\bibitem[Owocki et al. (1996)]{owcrga96}
         Owocki, S. P., Cranmer, S. R., Gayley, K. G. 
         1996, ApJ, 472, L115
%
\bibitem[Owocki et al. (1998)]{owgacr98}
         Owocki, S. P., Gayley, K. G., Cranmer, S. R. 1998
		 ASP Conf. Ser., 131, 237
%
\bibitem[Puls et al. (2008)]{puls+08}
         Puls, J., Vink, J. S., Najarro, F. 2008, A\&ARv, 16, 209
%
\bibitem[Shen \& Bildsten (2007)]{shen+07}
         Shen, K. J., \& Bildsten, L. 2007, ApJ, 660, 1444
%
\bibitem[Schr\"oder (1985)]{schroder85}
         Schr\"oder, K.-P. 1985, A\&A, 147, 103
%
\bibitem[Shore (2013)]{shore13}
         Shore, S. N. 2013, A\&A, 559, L7
%
\bibitem[Skopal (2005)]{sk05}
         Skopal, A. 2005, A\&A, 440, 995
%
\bibitem[Skopal (2015)]{sk15na2}
         Skopal, A. 2015, New A, in press [arXiv:1402.6126]
%
\bibitem[Skopal \& Carikov\'a ()]{skzc14}
         Skopal, A., Carikov\'a, Z. A\&A (submitted)
%
\bibitem[Skopal et al. (2014)]{sk+14}
         Skopal, A., Drechsel, H., Tarasova, T. N. et al. 2014, 
         A\&A, in press [arXiv:1407.8212]
%
\bibitem[Tutukov \& Yungelson (1976)]{tutukov+76}
         Tutukov, A. V., \& Yungelson, L. R. 1976, 
         Astrophysics, 12, 342
%
\bibitem[van Belle et al. (1999)]{bel+99}
         van Belle, G. T., Lane, B. F., Thompson, R. R., et al.
         1999, AJ, 117, 521
%
\bibitem[Zamanov et al. (2008)]{zam+08}
         Zamanov, R. K., Bode, M. F., Melo, C. H. F., et al. 
         2008, MNRAS, 390, 377
%
\end{thebibliography}
\end{document}